\documentclass[preprint,12pt,3p,times]{elsarticle}

\usepackage{amssymb}
\usepackage[numbers]{natbib}
 \usepackage{amsthm}

\journal{Nuclear Physics A}

\begin{document}

\begin{frontmatter}



\title{Quantum number assignments of light strange baryons in Regge phenomenology}

\author{Juhi Oudichhya$^{a}$}
\ead{juhioudichhya01234@gmail.com}



\affiliation[]{Department of Physics, Sardar Vallabhbhai National Institute of Technology, Surat, Gujarat-395007, India}

\author[]{Keval Gandhi$^{b}$}

\fnmark[2]

\ead{keval.physics@yahoo.com}



\affiliation[]{Department of Computer Sciences and Engineering,
	Institute of Advanced Research, Gandhinagar, Gujarat 382426, India}

\author[]{Ajay Kumar Rai$^{a}$}

\fnmark[2]

\ead{raiajayk@gmail.com}

\begin{abstract}
	The present article contains the descriptive study of light strange baryons $\Lambda$, $\Sigma$, $\Xi$, and $\Omega$. The Regge phenomenology with quasi-linear Regge trajectories has been employed and the relations between Regge slopes, intercepts, and baryon masses have been extracted. With the aid of these relations, ground state masses are obtained for $\Xi$ and $\Omega$ baryons. Further, the Regge parameters have also been estimated to calculate the excited state masses for $\Lambda$, $\Sigma$, $\Xi$, and $\Omega$ baryons in both the  $(J,M^{2})$ and $(n,M^{2})$ planes. The obtained results are compared with available experimental data and various theoretical approaches.  The spin-parity of recently observed light baryons are predicted in this work and our predictions may be useful in future experimental searches of these baryons and their $J^{P}$ assignments.  

\end{abstract}




\end{frontmatter}


\section{Introduction}\label{}

Hadron spectroscopy is a tool for understanding the dynamics of quark interactions in composite systems like mesons, baryons, and exotics. The spectroscopic study of baryon with a strange quark(s) will have a great interest because the strange quark would be slightly heavier than up ($u$) and down ($d$) quarks and considerably lighter than charm ($c$) and bottom ($b$) quarks. In this paper we present a description of hyperons which are $\Lambda$, $\Sigma$  baryons with strangeness -1, $\Xi$ and $\Omega$  baryons with strangeness -2 and -3, respectively. Many experimental facilities around the world have been seeking to study the strange baryons. Recently the ALICE Collaboration observed an attractive interaction between protons and $\Xi^{-}$ baryon \cite{ALICE2019}. In 2019 the BESIII Collaboration observed $\Xi(1530)$ in a baryon-antibaryon pair from charmonium decay \cite{BESIII2019}. At Jefferson Lab, the photoproduction of $\Xi$ has been observed using the CLAS detector \cite{Jefferson2005}. Furthermore, JLab has proposed to investigate the strange hyperon spectroscopy using a secondary KL beam and a GlueX experiment and the results are expected to provide more knowledge of strange hyperons $\Lambda$, $\Sigma$, and $\Xi$ \cite{Jlab2020}. Extensive research have also been conducted by the BABAR Collaboration \cite{BABAR2006}. Recently in 2018, the Belle Collaboration observed new excited $\Omega^{-}$ state as $\Omega(2012)$ through $e^{-}e^{+}$ annihilations decaying into $\Xi^{0}K^{-}$ and $\Xi^{-}\bar{K}^{0}$ channels \cite{Yelton2018}.

By studying the properties these, so-called, strange baryons or hyperons, hadron physicists will be able to make progress in answering various critical questions, namely; what is the intrinsic structure of these baryons? what are the important degrees of freedom in a baryon? and, are there exotic forms of baryon-like matter?. Eventually, addressing these questions helps in shedding light on the deeper fundamental question, namely “how to understand the underlying confinement mechanism?”. The anticipated multiplet structure of the baryons must be established experimentally to understand the symmetries and dynamics of the strong interaction, and the details of their excitation spectrum are vital for that. However, experimental information is presently very limited, in particular for $\Xi$ and $\Omega$ baryons with strangeness -2 and -3, respectively. For a large part, the lack of experimental knowledge can be understood by the fact that with the widely usage of electromagnetic probes the production cross section of strange baryons is very limited making it difficult to generate sufficient statistics. The $\Xi$ baryons are produced only at the final state of a decay process and have small cross-sections (typically a few $\mu b$) \cite{Pervin}. Somewhat this comment is also valid for $\Omega^-$ baryons. Also, $\Omega^-$ baryons have zero isospins, which means that $\Omega^{*-} \rightarrow \Omega^- \pi^0$ decays are highly suppressed, and this restricts the possible decays of excited states. Therefore, $\Xi K$ being the excepted decay mode for the low-lying $\Omega^-$ states  \cite{Yelton2018}. These decays are similar to the $\Omega_c^0 \rightarrow \Xi_c^+ K^-$ decays discovered by the LHCb \cite{Aaij2017} Collaboration and confirmed by the Belle \cite{Yelton2018_2} Collaboration shortly after. 

\begin{table}
	\centering
	\caption{Masses and $J^{P}$ values of $\Xi$ and $\Omega$ baryons are listed in PDG \cite{Zyla2020}. The status is given as poor(*), fair(**), very likely(***), and certain(****).}\label{tab:table1}
	\begin{tabular*}{80mm}{@{}llll@{}}
		\hline	\noalign{\smallskip}
		Resonance  & Mass (MeV) & $J^{P}$ & Status \\ 
		\hline\noalign{\smallskip}
		$\Xi^{0}$ &  1314.86 $\pm$ 0.20 & $\frac{1}{2}^{+}$ & ****\\
		$\Xi^{-}$ & 1321.71 $\pm$ 0.07 & $\frac{1}{2}^{+}$ & ****\\
		$\Xi(1530)^{0}$ & 1531.80 $\pm$ 0.32 & $\frac{3}{2}^{+}$ & **** \\
		$\Xi(1530)^{-}$ & 1535.0 $\pm$ 0.6 & $\frac{3}{2}^{+}$ & **** \\
		$\Xi(1620)$ & 1620 &  & *\\
		$\Xi(1690)$ & 1690 $\pm$ 10 &  & ***\\
		$\Xi(1820)$   & 1823 $\pm$ 5 & $\frac{3}{2}^{-}$ & ***\\
		$\Xi(1950)$ & 1950 $\pm$ 15 &  & ***\\
		$\Xi(2030)$  & 2025 $\pm$ 5 & $\frac{5}{2}^{?}$ & ***\\
		$\Xi(2120)$  & 2120  &  & *\\
		$\Xi(2250)$ & 2250 & & **\\
		$\Xi(2370)$ & 2370 &  & **\\
		$\Xi(2500)$ & 2500 & & *\\
		
		$\Omega^{-}$  & 1672.45 $\pm$ 0.29 & $\frac{3}{2}^{+}$  & ****\\
		$\Omega(2012)^{-}$ & 2012.4 $\pm$ 0.9 &  & ***\\
		$\Omega(2250)^{-}$ & 2252 $\pm$ 9 &  & ***\\
		$\Omega(2380)^{-}$ & 2380  &  & **\\
		$\Omega(2470)^{-}$ &2474 $\pm$ 12 & & **\\
		\hline
	\end{tabular*}
\end{table}	

$\overline{\mbox{\sffamily P}}${\sffamily ANDA} (antiProton ANnihilations at DArmstadt), the upcoming experimental facility at FAIR (Facility for AntiProton and Ion Research), will have a task to establish the whole spectrum of hyperons through antiproton-proton annihilation \cite{Panda1,Panda2,Panda3,Panda4,Panda6,Barucca2021_1,Panda5,Abazov}. Recently, a member of the $\overline{\mbox{\sffamily P}}${\sffamily ANDA} group studied the feasibility of the reaction $\bar{p}p \rightarrow \bar{\Xi}^+ \Xi^{*-}$ and its charge conjugate channel, where the $\Xi^*$ denotes the following intermediate resonance states: $\Xi$ (1530), $\Xi$ (1690), and $\Xi$ (1820) \cite{Barucca2021_2}. A major goal of the $\Xi$ spectroscopy program at $\overline{\mbox{\sffamily P}}${\sffamily ANDA} is the determination of the spin and parity quantum numbers of the $\Xi$ states \cite{Putz2020}. Unlike $\Xi$ and $\Omega$, $\Lambda$ and $\Sigma$ baryons have quite a number of experimentally established states. Table \ref{tab:table1} represents the excited $\Xi$ and $\Omega$ baryons listed by the latest version of the Particle Data Group (PDG) \cite{Zyla2020}. Spin and parity quantum numbers of newly observed $\Xi$ and $\Omega$ states are not yet confirmed and the PDG needs more confirmations. It is crucial to assign the spin-parity ($J^{P}$) of hadrons which facilitate the determination of properties such as decay width, branching fraction, isospin mass splitting, polarization amplitude, etc.
Accordingly, we have studied the light baryons which are made of the quarks $u$, $d$, and $s$ only. The SU(3) flavour symmetry can be only approximate because the mass of the strange quark is about 0.1 GeV greater than the masses of the up and down quarks, although this mass difference is relatively small compared to the typical QCD binding energy which is of order 1 GeV.
The eightfold way and the standard SU(3) Gell-Mann–Okubo (GMO) formula \cite{Gell-Mann} have played an important role in particle physics. However, the direct generalization of the GMO formula to the charmed and bottom hadrons cannot agree well with experimental data due to higher-order breaking effects. However, for light baryons the breaking of SU(3) symmetry is minimum.

Several phenomenological and theoretical model have been employed to study the properties of light baryons. Recently the review articles \cite{Hua-Xing2023,F. Gross2023},  provides a complete analysis of the recent experimental and theoretical progress in the field of hadron physics. The authors of Ref.\cite{Faustov} estimated the mass spectra of strange baryons using the relativistic quark model based on the quasi-potential approach and baryons are treated as relativistic quark-diquark bound systems.  In the recent study \cite{Menapara,ChandniIJMPA,ChandniIJMPA2023,ChandniOmega,zalak}, the authors employed hypercentral Constituent Quark Model with linear confining potential along with first order correction term to obtain the mass spectra of light baryons and also light strange baryons with strangeness -1,-2, and -3. Regge trajectories are explored for the linearity of the calculated masses for $(J, M^2)$ and $(n, M^2)$. Ref. \cite{Metsch} present the complete excited $\Lambda$, $\Sigma$, $\Xi$, and $\Omega$ spectrum in a relativistic quark model based on the three quark Bethe-Salpeter equation with instantaneous two and three body forces. The investigation of instanton-induced effects in the baryon mass spectrum plays a central role in this work. In a recent study \cite{Xiao2021},the nature of two poles for the $\Lambda(1405)$ state is discussed in which they revisit the ineteraction of $\bar{K}N$ and $\pi\Sigma$ with their coupled channels, where two pole structure is found. They also generate the two poles in the single channel interaction of $\bar{K}N$ and $\pi\Sigma$, respectively. From their results they conclude that the $\Lambda(1405)$ state may be overlapped with two different states of the same quantum numbers. 
The authors of Ref. \cite{Bijker2000} present a systematic analysis of spectra and transition rates of strange baryons within the framework of a collective string-like $qqq$ model in which the orbital excitations are treated as rotations and vibrations of the strings.  The study \cite{ELSA} discussed the recent results and the ongoing new BGO-OD experiment at ELSA.

In our previous work we have used the Regge phenomenology with the assumption of linear Regge trajectories to obtain the mass spectra of singly, doubly, and triply heavy baryons \cite{Oudichhya,Oudichhya2,physica,Juhiarxiv}. 
In the present article, we give a systematic study of light strange baryons $\Lambda$, $\Sigma$, $\Xi$, and $\Omega$ by employing the Regge theory.
We find the relations between the intercept, slope ratios, and baryon masses in both the $(J, M^2)$ and $(n, M^2)$ planes \cite{Wei2008}. With the help of these relations, we determine the ground state masses of $\Xi$ and $\Omega$ baryons. We extract the Regge parameters (a(0)and $\alpha^{,}$) to determine the mass spectra of all light strange baryons in both the $(J, M^2)$ and $(n, M^2)$ planes. It is evident that the ground and low-lying resonance states are within a reasonable range for almost all of the models and approaches. However, the higher excited states exhibit huge variations in their mass predictions, which motivated us to study the experimental determination of the spin and parity quantum numbers of these light strange baryons. 

The remainder of this paper is organized as follows. After briefing various experimental and theoretical approaches, in Sec. 2 we describe Regge theory and  extract the mass relations. By using these relations we obtain the masses for $\Lambda$, $\Sigma$, $\Xi$, and $\Omega$ baryons in the $(J,M^{2})$ plane for natural and unnatural parity states. Further we obtain the Regge parameters for each Regge line and calculate the radial and orbital excited states of these baryons in the $(n,M^{2})$ plane. We extend this model and try to determine the  remaining states other than natural and unnatural parity states in the $(J,M^{2})$ plane. The detailed description of our results obtained is discussed in Sec. 3. Finally we concluded our study in Sec. 4.

\section{Theoretical Framework}

To study the hadron spectroscopy, the linear Regge trajectory is one of the most effective and widely used phenomenological approach. The plots of Regge trajectories of hadrons in the $(J,M^{2})$ plane are usually called Chew-frautschi plots \cite{Chew1961}. They used the theory to study the strong quark gluon interaction and observed that the excited states of experimentally missing mesons and baryons exists on the linear trajectories in the  $(J,M^{2})$ plane. The trajectory of a particular pole is characterized by a set of internal quantum numbers and the hadrons lying on the particular Regge line have the same internal quantum numbers. The most general form of linear Regge trajectories can be expressed as \cite{Nambu1974,Nambu1979,Oudichhya,Wei2008},
\begin{eqnarray}
	\label{eq:1}
	J = \alpha(M) = a(0)+\alpha^{'} M^{2},
\end{eqnarray}	
where $a(0)$ and $\alpha^{'}$ are, respectively, the intercept and slope of the Regge trajectory. Two relations  were derived in a $q\overline{q}$ string picture of hadrons \cite{Add4} and are generalised for baryons as well \cite{Add3,Add6,Approx}, which are expressed as (see \cite{Wei2008} and references therein). 
\begin{equation}
	\label{eq:2}
	a_{iiq}(0) + a_{jjq}(0) = 2a_{ijq}(0) ,	
\end{equation}
\begin{eqnarray}
	\label{eq:3}
	\frac{1}{{\alpha^{'}}_{iiq}} + \frac{1}{{\alpha^{'}}_{jjq}} = \frac{2}{{\alpha^{'}}_{ijq}} ,	
\end{eqnarray}
where $i, j, q$ represent the quark flavors. The Ref. \cite{Add3} showed that the additivity of inverse Regge
slopes is consistent with the formal chiral and heavy quark
limits for both mesons and baryons.

Now Using Eqs. (\ref{eq:1}) and (\ref{eq:2}) we obtain,
\begin{equation}
	\label{eq:4}
	\alpha^{'}_{iiq}M^{2}_{iiq}+\alpha^{'}_{jjq}M^{2}_{jjq}=2\alpha^{'}_{ijq}M^{2}_{ijq} .
\end{equation}
We get two pairs of solutions after combining the relations (\ref{eq:3}) and (\ref{eq:4}) which are expressed as \cite{Wei2008},
\begin{eqnarray}
	\label{eq:5}
	\frac{\alpha^{'}_{jjq}}{\alpha^{'}_{iiq}}=\frac{1}{2M^{2}_{jjq}}\times[(4M^{2}_{ijq}-M^{2}_{iiq}-M^{2}_{jjq}) 
	\pm\sqrt{{{(4M^{2}_{ijq}-M^{2}_{iiq}-M^{2}_{jjq}})^2}-4M^{2}_{iiq}M^{2}_{jjq}}],
\end{eqnarray}
and, 
\begin{eqnarray}
	\label{eq:6}	
	\frac{\alpha^{'}_{ijq}}{\alpha^{'}_{iiq}}=\frac{1}{4M^{2}_{ijq}}\times[(4M^{2}_{ijq}+M^{2}_{iiq}-M^{2}_{jjq}) 
	\pm\sqrt{{{(4M^{2}_{ijq}-M^{2}_{iiq}-M^{2}_{jjq}})^2}-4M^{2}_{iiq}M^{2}_{jjq}}].
\end{eqnarray}
These are the significant relationships that we have obtained between slope ratios and baryon masses.
Now Eq. (\ref{eq:5}) obtained above can also be expressed as,

\begin{equation}
	\label{eq:7}
	\frac{\alpha^{'}_{jjq}}{\alpha^{'}_{iiq}}=	\frac{\alpha^{'}_{kkq}}{\alpha^{'}_{iiq}}\times	\frac{\alpha^{'}_{jjq}}{\alpha^{'}_{kkq}} ,
\end{equation}	
here $k$ can be any quark flavor. Thus we have,

\begin{eqnarray}
	\label{eq:8}
	\frac{[(4M^{2}_{ijq}-M^{2}_{iiq}-M^{2}_{jjq})+\sqrt{{{(4M^{2}_{ijq}-M^{2}_{iiq}-M^{2}_{jjq}})^2}-4M^{2}_{iiq}M^{2}_{jjq}}]}{2M^{2}_{jjq}} \\ \nonumber
	=\frac{[(4M^{2}_{ikq}-M^{2}_{iiq}-M^{2}_{kkq})+\sqrt{{{(4M^{2}_{ikq}-M^{2}_{iiq}-M^{2}_{kkq}})^2}-4M^{2}_{iiq}M^{2}_{kkq}}]/2M^{2}_{kkq}}{[(4M^{2}_{jkq}-M^{2}_{jjq}-M^{2}_{kkq})+\sqrt{{{(4M^{2}_{jkq}-M^{2}_{jjq}-M^{2}_{kkq}})^2}-4M^{2}_{jjq}M^{2}_{kkq}}]/2M^{2}_{kkq}} .
\end{eqnarray}

This is the general relation in terms of baryon masses which can be used to estimate the mass of any baryon state if all other masses are known. In the present work, the ground-state ($J^{P}=\frac{1}{2}^{+}$ and $\frac{3}{2}^{+}$) masses of the light strange baryons having strangeness -2 and -3 are evaluated using the above equation.	

\subsection{Masses in the $(J,M^{2})$ plane}
In this section using the relations we have extracted above, we determine the ground-state masses of $\Xi$ and $\Omega$ baryons, as well as the orbitally excited state masses of all the light strange baryons $\Lambda$, $\Sigma$, $\Xi$, and $\Omega$ for natural ($P=(-1)^{J-\frac{1}{2}}$) and unnatural ($P=(-1)^{J+\frac{1}{2}}$) parities in the $(J,M^{2})$ plane. The quark combination of $\Xi^{0}$ baryon is $uss$, hence we put $i=d$, $j=s$, $q=u$, and $k=d$ in Eq. (\ref{eq:8}). We obtain the mass expression for $\Xi^{0}$ as a function of squared masses of neutron ($n$) and $\Lambda^{0}$ baryons, which is expressed as

\begin{eqnarray}
	\label{eq:9} 
	\left[(M_{n}+M_{\Xi^{0}})^{2} -4M^{2}_{\Lambda^{0}}\right] 
	= \sqrt{(4M^{2}_{\Lambda^{0}}-M^{2}_{n}-M^{2}_{\Xi^{0}})^{2} - 4M^{2}_{n}M^{2}_{\Xi^{0}}} ;
\end{eqnarray}
substituting the masses of $n$ and $\Lambda^{0}$ into Eq. (\ref{eq:9}) from PDG \cite{Zyla2020}, we get the ground-state mass of $\Xi^{0}$ as 1291.80 MeV for $J^{P}=\frac{1}{2}^{+}$. Similarly we can get 1534.38 for $J^{P}=\frac{3}{2}^{+}$. 
Since we have taken the experimental masses as inputs, hence the experimental error has been incorporated in the present work. We expressed our obtained mass as, $Mass = M \pm \delta M$, where $\delta M$ is the experimental error in the mass. Hence, after introducing the error propogation in the above equation we get,

\begin{equation}
	\frac{\delta
	M_{\Xi}}{M_{\Xi}}=\frac{\sqrt{(32M_{\Lambda}\delta M_{\Lambda})^{2}+(2M_{n}\delta M_{n})^{2}+16M^{2}_{n}M^{2}_{\Xi}\left[ \left(\frac{\delta M_{n}}{M_{n}}\right)^{2}+\left(\frac{\delta M_{\Xi}}{M_{\Xi}}\right)^{2}\right]}}{8M^{2}_{\Lambda}-2M^{2}_{n}-4M_{n}M_{\Xi}}
\end{equation}
putting the values $M_{n}$ = 939.56 MeV, $\delta M_{n}$= 0.0000005 MeV, $M_{\Lambda}$ = 1115.68 MeV, $\delta M_{\Lambda}$ = 0.006 MeV for $J^{P}= \frac{1}{2}^{+}$ state from PDG \cite{Zyla2020} and $M_{\Xi^{0}}$ = 1291.80 MeV (calculated above) into Eq. (10), we get $\delta M_{\Xi^{0}}$ 0.012 MeV. Hence, we get ground state mass of $\Xi^{0}$ baryon as 1291.80$\pm$0.012 MeV. 
In the same manner we can yield for $\frac{3}{2}^{+}$ state (see Table \ref{tab:table4}). 

Similarly to evaluate the ground state ($J^{P}=\frac{3}{2}^{+}$) mass of $\Omega^{-}$ baryon, composed of three strange quarks ($sss$), we put  $i=u$, $j=s$, $q=s$, and $k=u$ in Eq. (\ref{eq:8}) we get, 
\begin{eqnarray}
	\label{eq:10} 
	\left[(M_{\Sigma^{*+}}+M_{\Omega^{-}})^{2} - 4M^{2}_{\Xi^{*0}}\right]  
	=\sqrt{(4M^{2}_{\Xi^{*0}}-M^{2}_{\Sigma^{*+}}-M^{2}_{\Omega^{-}})^{2}-4M^{2}_{\Sigma^{*+}}M^{2}_{\Omega^{-}}};
\end{eqnarray}
and the expression to obtained the experimental error in the calculated mass for $\Omega$ baryon is given as,

\begin{equation}
	\frac{\delta
		M_{\Omega}}{M_{\Omega}}=\frac{\sqrt{(32M_{\Xi}\delta M_{\Xi})^{2}+(2M_{\Sigma}\delta M_{\Sigma})^{2}+16M^{2}_{\Sigma}M^{2}_{\Omega}\left[ \left(\frac{\delta M_{\Sigma}}{M_{\Sigma}}\right)^{2}+\left(\frac{\delta M_{\Omega}}{M_{\Omega}}\right)^{2}\right]}}{8M^{2}_{\Xi}-2M^{2}_{\Sigma}-4M_{\Sigma}M_{\Omega}}
\end{equation}

Hence, substituting the masses of $\Sigma^{*+}$ \cite{Zyla2020} and $\Xi^{*0}$ (calculated above) into the above equations, we obtain the ground-state mass of $\Omega^{-}$ baryon as 1685.93$\pm$1.85 MeV. 	
Now, after evaluating the ground-state masses, the orbitally excited states of light strange baryons are calculated lying on $\frac{1}{2}^{+}$ and $\frac{3}{2}^{+}$ trajectories by obtaining the Regge slopes $\alpha^{'}$. For instance using Eq. (\ref{eq:5}), 

\begin{eqnarray} 
	\label{eq:11} 
	\frac{\alpha^{'}_{\Xi^{0}}}{\alpha^{'}_{N}}=\frac{1}{2M^{2}_{\Xi^{0}}}\times[(4M^{2}_{\Lambda^{0}}-M^{2}_{n}-M^{2}_{\Xi^{0}}) 
	+\sqrt{{{(4M^{2}_{\Lambda^{0}}-M^{2}_{n}-M^{2}_{\Xi^{0}}})^2}-4M^{2}_{n}M^{2}_{\Xi^{0}}}],
\end{eqnarray}
inserting the masses of $\Xi^{0}$, $n$, and $\Lambda^{0}$ baryons in the above equation we obtain $\alpha^{'}_{\Xi^{0}}/\alpha^{'}_{N}$. Here also, we have taken the experimental masses as inputs to evaluate the Regge slopes, hence the experimental error is incorporated. Here, we expressed our estimated Regge slopes as, $\alpha^{'} \pm \delta \alpha^{'}$, where $\delta \alpha^{'}$ is the error in the slope.

Now, from Eq. (\ref{eq:1}) we can have $\alpha^{'}_{N} = 2/(M^{2}_{N(1680)}-M^{2}_{n})$. Hence, we can calculate $\alpha^{'}_{\Xi^{0}}$ = 0.7526$\pm$0.0014 GeV$^{-2}$ for $\frac{1}{2}^{+}$ trajectory. Similarly with Eqs. (\ref{eq:5}) and (\ref{eq:6}) we can determine the Regge slopes of $\Lambda$, $\Sigma$,  and $\Omega$ baryons for both $\frac{1}{2}^{+}$ and $\frac{3}{2}^{+}$ trajectories. According to the Ref \cite{Approx}. $\alpha^{'}_{\Lambda}$ $\simeq$ $\alpha^{'}_{\Sigma}$, in this work we take this approximation.
\begin{table}
	\caption{Regge slopes of $\frac{1}{2}^{+}$ and $\frac{3}{2}^{+}$ trajectories of the light strange baryons. (in GeV$^{-2}$).}\label{tab:tablea}
	\begin{tabular*}{150mm}{@{}ccccccccccc@{}}
			\hline	\noalign{\smallskip}
		&$\Lambda$  &$\Sigma$&$\Xi$  &
		$\Omega$  \\
		\hline
	\noalign{\smallskip}
		$\alpha^{'} \pm \delta \alpha^{'}$ (S = 1/2)  & 0.8609$\pm$0.001 &0.8609$\pm$0.001 &0.7526$\pm$0.001 & - \\
		$\alpha^{'*} \pm \delta \alpha{'*}$ (S = 3/2) & - &	0.8189$\pm$0.053 &	0.7370$\pm$0.098&	0.6763$\pm$0.099 \\
		\hline
	\noalign{\smallskip}
	\end{tabular*}
\end{table}	
Table \ref{tab:tablea} shows the estimated values of $\alpha^{'} \pm \delta \alpha^{'}$ and $\alpha^{'*} \pm \delta \alpha^{'}$ of the light strange baryons.
Now from Eq. (\ref{eq:1}) one can write,

\begin{equation}
	\label{eq:12}
	M_{J+1} = \sqrt{M_{J}^{2}+\frac{1}{\alpha^{'}}} ,
\end{equation}
With the help of values of Regge slopes extracted for the light-strange baryons, from Eq.(\ref{eq:12}), the masses of orbitally excited states with $J^{P}= \frac{3}{2}^{-}, \frac{5}{2}^{+}, \frac{7}{2}^{-}$.... and $J^{P}= \frac{5}{2}^{-}, \frac{7}{2}^{+}, \frac{9}{2}^{-}$.... for natural and unnatural parities respectively, in the $(J,M^{2})$ plane can be evaluated. The numerical results are shown in Tables \ref{tab:table2}-\ref{tab:table5}. Here the spectroscopic notations \textit{$N^{2S+1}L_{J}$} are used to represent the state of the particles, where $N$, $L$, $S$ denote the radial excited quantum number, orbital quantum number, and intrinsic spin, respectively.

\begin{table}
	\centering
	\caption{Masses of excited states of $\Lambda$ baryon in the $(J,M^{2})$ plane with natural parities. The numbers in the boldface are the experimental values taken as the input \cite{Zyla2020} (in MeV).}\label{tab:table2}
	\begin{tabular*}{148mm}{c@{\extracolsep{\fill}}cccccccccccccccc}
		\hline
		States&\multicolumn{1}{c}{Present}& & &Others \\
		\hline
		\textit{$N^{2S+1}L_{J}$}&$\Lambda^{0}$ & PDG \cite{Zyla2020} & \cite{Faustov} & \cite{Capstick}  &\cite{Menapara} &\cite{Metsch} & \cite{Santopinto2015} \\	
		\hline
		\noalign{\smallskip}
		$1^{2}S_{\frac{1}{2}}$ & \textbf{1115.68$\pm$0.006} &1115.68$\pm$0.006 &1115 &1115 &1115 &1108 &1116\\
		$1^{2}P_{\frac{3}{2}}$ & 1551.23$\pm$0.43 &1519 &1549 &1545 &1534 &1508 &1650\\
		$1^{2}D_{\frac{5}{2}}$ & 1888.89$\pm$0.50 &1820 &1825 &1890 &1746 &1834 &1896\\
		$1^{2}F_{\frac{7}{2}}$ & 2174.73$\pm$0.53 &2100 &2097 &2150 &1970 &2090\\
		$1^{2}G_{\frac{9}{2}}$ & 2427.15$\pm$0.55 &2350 &2360 & & 2204 &2340\\
		$1^{2}H_{\frac{11}{2}}$& 2655.68$\pm$0.56 & &2605\\
	\hline
	\end{tabular*}
\end{table}

\begin{table}[h!]
	\caption{Masses of excited states of $\Sigma$ baryon in the $(J,M^{2})$ plane with natural  and unnatural parities. The numbers in the boldface are the experimental values taken as the inputs \cite{Zyla2020} (in MeV).}\label{tab:table3}
	\begin{tabular*}{175mm}{c@{\extracolsep{\fill}}cccccccccccccccc}
		\hline\noalign{\smallskip}
		States&\multicolumn{3}{c}{Present}& & &Others	\\
		\hline\noalign{\smallskip}
		\textit{$N^{2S+1}L_{J}$}&$\Sigma^{+}$ &$\Sigma^{0}$ &$\Sigma^{-}$ & PDG \cite{Zyla2020} & \cite{Faustov} & \cite{Capstick}  &\cite{Menapara} &\cite{Metsch} \\
		\hline\noalign{\smallskip}
		$1^{2}S_{\frac{1}{2}}$ &\textbf{1189.37$\pm$0.07} &\textbf{1192.64$\pm$0.024} &\textbf{1197.45$\pm$0.03} & &1187 &1190 &1193 &1190  \\
		$1^{2}P_{\frac{3}{2}}$ &1605.05$\pm$0.42 &1607.47$\pm$0.42 &1611.04$\pm$0.41 &1675 & 1706 &1655 &1698 &1669\\
		$1^{2}D_{\frac{5}{2}}$ &1933.33$\pm$0.49 &1935.34$\pm$0.49 &1938.31$\pm$0.49 &1915 &1991 &1995 &1998 &1956\\
		$1^{2}F_{\frac{7}{2}}$ &2213.44$\pm$0.52 &2215.20$\pm$0.52 &2217.79$\pm$0.52 & &2259 &2245 &2318 &2236 \\
		$1^{2}G_{\frac{9}{2}}$ &2461.89$\pm$0.54 &2463.47$\pm$0.54 &2465.80$\pm$0.54 & &2548 \\
		$1^{2}H_{\frac{11}{2}}$& 2687.47$\pm$0.55 &2688.92$\pm$0.55 &2691.05$\pm$0.55\\
		\noalign{\smallskip}\noalign{\smallskip}
		$1^{4}S_{\frac{3}{2}}$ & \textbf{1382.83$\pm$0.34} &\textbf{1383.70$\pm$1.00} &\textbf{1387.20$\pm$0.50} & &1381 &1370 &1384 &1411 \\
		$1^{4}P_{\frac{5}{2}}$ & 1770.13$\pm$17.52 &1770.81$\pm$17.53 &1773.55$\pm$17.49 &1775 &1757 &1755 &1680 &1770\\
		$1^{4}D_{\frac{7}{2}}$ & 2086.75$\pm$21.02 &2087.33$\pm$21.02 &2089.65$\pm$20.99 &2030 &2033 &2060 &1962 &2070\\
		$1^{4}F_{\frac{9}{2}}$ & 2361.28$\pm$22.75 &2361.80$\pm$22.75 &2363.85$\pm$22.72 & &2289 & &2257 &2325\\
		$1^{4}G_{\frac{11}{2}}$& 2607.07$\pm$23.79 &2607.53$\pm$23.79 &2609.39$\pm$23.77 &2620 &2529\\
		$1^{4}H_{\frac{13}{2}}$& 2831.60$\pm$24.49 &2832.03$\pm$24.49 &2833.74$\pm$24.97\\	
	\hline
	\end{tabular*}
\end{table}

\begin{table}
	T\centering
	\caption{Masses of excited states of $\Xi$ baryon in the $(J,M^{2})$ plane with natural  and unnatural parities. (in MeV).}\label{tab:table4}
	\begin{tabular*}{175mm}{c@{\extracolsep{\fill}}cccccccccccccccc}
		\hline\noalign{\smallskip}
		States&\multicolumn{2}{c}{Present}& & &Others	\\
		\hline\noalign{\smallskip}
		\textit{$N^{2S+1}L_{J}$}&$\Xi^{0}$ &$\Xi^{-}$ & PDG \cite{Zyla2020} & \cite{Faustov} & \cite{Capstick} & \cite{Metsch} &\cite{Menapara} &\cite{Y. Chen} \\	
		\hline
		\noalign{\smallskip}
		$1^{2}S_{\frac{1}{2}}$ & 1291.80$\pm$0.01 &1293.06$\pm$0.01 &1314.80$\pm$0.20 ($\Xi^{0}$) &1330 &1305 & 1310 &1322 &1317 \\
		& & &1321.71$\pm$0.07 ($\Xi^{-}$)  \\
		$1^{2}P_{\frac{3}{2}}$ & 1731.32$\pm$0.71 &1732.26$\pm$0.71 & &1764 &1785 &1780 &1871 &1801 \\
		$1^{2}D_{\frac{5}{2}}$ & 2079.95$\pm$0.83 &2080.74$\pm$0.84 &2025$\pm$5 &2108 &2045 &2013 &2234 &1959 &\\
		$1^{2}F_{\frac{7}{2}}$ & 2378.01$\pm$0.89 &2378.69$\pm$0.90 &2370 &2460 &2355 &2320 &2647\\
		$1^{2}G_{\frac{9}{2}}$ & 2642.66$\pm$0.93 &2643.28$\pm$0.93 & &    & &\\
		$1^{2}H_{\frac{11}{2}}$& 2883.12$\pm$0.95 &2883.69$\pm$0.95 \\
		\noalign{\smallskip}\noalign{\smallskip}
		$1^{4}S_{\frac{3}{2}}$ & 1534.38$\pm$0.91 &1530.76$\pm$1.55 & 1531.80$\pm$0.32 ($\Xi^{*0}$) &1518 &1505 &1539 &1531 &1526 \\
		& & & 1535.0$\pm$0.6 ($\Xi^{*-}$)\\
		$1^{4}P_{\frac{5}{2}}$ & 1922.61$\pm$31.21 &1919.72$\pm$31.27 &1950$\pm$15 &1853 &1900 &1955 &1859 &1917\\
		$1^{4}D_{\frac{7}{2}}$ & 2244.67$\pm$37.79 &2242.19$\pm$37.85 & 2250 &2189 &2180 &2169 &2203 &2074\\
		$1^{4}F_{\frac{9}{2}}$ & 2525.99$\pm$41.13 &2523.79$\pm$41.18 & &2502 & & 2505 &2588\\
		$1^{4}G_{\frac{11}{2}}$& 2778.98$\pm$43.17 &2776.98$\pm$43.21\\
		$1^{4}H_{\frac{13}{2}}$& 3010.79$\pm$44.55 &3008.94$\pm$44.58\\
	\hline
	\end{tabular*}
\end{table}

\begin{table}
	\centering
	\caption{Masses of excited states of $\Omega$ baryon in the $(J,M^{2})$ plane with unnatural parities (in MeV).}\label{tab:table5}
	\begin{tabular*}{120mm}{c@{\extracolsep{\fill}}cccccccccccccccc}
		\hline\noalign{\smallskip}
		States & Present & PDG \cite{Zyla2020}& \cite{Faustov} &\cite{Capstick} & \cite{Metsch} & \cite{ChandniOmega} 	\\
		\textit{$N^{2S+1}L_{J}$}\\
		\hline\noalign{\smallskip}
		
		\noalign{\smallskip}
	$1^{4}S_{\frac{3}{2}}$ & 1685.93$\pm$1.85 &1672.45$\pm$0.29 &1678 &1635 &1673\\
		$1^{4}P_{\frac{5}{2}}$ & 2077.97$\pm$38.27 & &2653 &2490 &2528 &1970\\
		$1^{4}D_{\frac{7}{2}}$ & 2406.97$\pm$46.70 &2380 &2369 &2295 &2292&2233 \\
		$1^{4}F_{\frac{9}{2}}$ & 2696.13$\pm$51.06 & &2649 & &2606&2521\\
		$1^{4}G_{\frac{11}{2}}$& 2957.14$\pm$53.75\\
		$1^{4}H_{\frac{13}{2}}$& 3196.91$\pm$55.58\\		
		
	\hline
	\end{tabular*}
\end{table}

\subsection{Masses in the $(n,M^{2})$ plane}

In this section the masses for radial excited states from  $1^{2}S_{\frac{1}{2}}-6^{2}S_{\frac{1}{2}}, 1^{4}S_{\frac{3}{2}}-6^{4}S_{\frac{3}{2}}, 1^{2}P_{\frac{3}{2}}-5^{2}P_{\frac{3}{2}}, 1^{4}P_{\frac{5}{2}}-5^{4}P_{\frac{5}{2}}, $
$1^{2}D_{\frac{5}{2}}-5^{2}D_{\frac{5}{2}}$   and
$1^{4}D_{\frac{7}{2}}-5^{4}D_{\frac{7}{2}}$  are estimated in the $(n,M^{2})$ plane.  The general equation for linear Regge trajectories in the ($n,M^{2}$) plane can be expressed as,
\begin{equation}
	\label{eq:13}
	n = \beta_{0} + \beta M^{2},
\end{equation} 

where $n$ = 1, 2, 3.... is the radial principal quantum number, $\beta_{0}$, and $\beta$ are the Regge intercept and slope of the trajectories. These parameters are extracted for each Regge line for $\Lambda$, $\Sigma$, $\Xi$, and $\Omega$ baryons. Since, the baryon multiplets lying on the single Regge line have the same Regge slope ($\beta$) and Regge intercept ($\beta_{0}$). Using relation (\ref{eq:13}) and the values for $\beta_{0}$ and $\beta$ obtained, we estimated the excited state masses of light strange baryons ($\Lambda$, $ \Sigma$, $\Xi$, and $\Omega$) lying on each Regge lines for natural and unnatural parity states.
For instance, using the slope equation, we have $\beta_{(S)} = 1/(M^{2}_{\Xi(2S)}-M^{2}_{\Xi(1S)})$ for $\Xi^{0}$ baryon, where $M_{\Xi(1S)}$ = 1291.80$\pm$0.01 MeV (calculated above) and taking $M_{\Xi(2S)}$ = 1886 MeV from \cite{Faustov} for the $1/2^{+}$ trajectory, we get $\beta_{(S)}$ = 0.52959$\pm$0.724$\times$10$^{-5}$ GeV$^{-2}$. From Eq. (\ref{eq:13}) we can write,

\begin{eqnarray}
	\label{eq:14}
	1 &=& \beta_{0(S)} + \beta_{(S)} M^{2}_{\Xi(1S)},\\ \nonumber
	2 &=& \beta_{0(S)} + \beta_{(S)} M^{2}_{\Xi(2S)},
\end{eqnarray}
using the above relations, we get $\beta_{0(S)}$ = 0.11024$\pm$0.00002. Here also we have calculated the error in Regge parameters. With the help of $\beta_{(S)}$ and $\beta_{0(S)}$, we calculate the masses of the excited $\Xi^{0}$ baryon for $n$ = 3, 4, 5... Similarly, we can express these relations for $P$ and $D$-wave as,

\begin{eqnarray}
	\label{eq:15} \nonumber
	1 &=& \beta_{0(P)} + \beta_{(P)} M^{2}_{\Xi(1P)},\\  \nonumber
	2 &=& \beta_{0(P)} + \beta_{(P)} M^{2}_{\Xi(2P)},\\
	1 &=& \beta_{0(D)} + \beta_{(D)} M^{2}_{\Xi(1D)},\\ \nonumber
	2 &=& \beta_{0(D)} + \beta_{(D)} M^{2}_{\Xi(2D)},
\end{eqnarray}

\begin{table}[h!]
	\caption{The values for Regge intercepts ($\beta_{0}$) and Regge slopes ($\beta$) (in GeV$^{-2}$) of light strange baryons in the $(n,M^{2})$ plane.}\label{tab:tableb}
	\begin{tabular*}{170mm}{c@{\extracolsep{\fill}}cccccccccccccccc}
		\hline\noalign{\smallskip}
		Trajectories &  $\beta \pm \delta \beta$ &$\beta_{0} \pm \delta \beta_{0} $ &  $\beta \pm \delta \beta$ &$\beta_{0} \pm \delta \beta_{0} $ \\	
		\hline\noalign{\smallskip}
		& \multicolumn{2}{c}{$\Lambda^{0}$}  & \multicolumn{2}{c}{$\Sigma^{+}$} \\
		\hline
		($\frac{1}{2}^{+}$, S) & 0.76031$\pm$0.773$\times$10$^{-5}$ &0.05361$\pm$0.00001 &0.66097$\pm$0.00007 &0.06498$\pm$0.00015 \\
		($\frac{3}{2}^{+}$, S) &-&-&0.64315$\pm$0.00039&-0.22986$\pm$0.00096  \\
		
		\noalign{\smallskip}
		($\frac{3}{2}^{-}$, P) &2.22328$\pm$0.00659 & -4.34992$\pm$0.01585 & 0.46416$\pm$0.00029 &-0.19576$\pm$0.00097 \\
		($\frac{5}{2}^{-}$, P) &-&- &0.56547$\pm$0.01983 &-0.77182$\pm$0.01736\\
		
		\noalign{\smallskip}
		($\frac{5}{2}^{+}$, D) &1.24969$\pm$0.00294 & -3.45879$\pm$0.01075 &0.43310$\pm$0.00035 &-0.61884$\pm$0.00156\\
		($\frac{7}{2}^{+}$, D) & - &- &0.57261$\pm$0.02876 &-1.49346$\pm$0.13493\\
		
		\hline\noalign{\smallskip}
		& \multicolumn{2}{c}{$\Xi^{0}$}  & \multicolumn{2}{c}{$\Omega^{-}$} \\
		\hline
		($\frac{1}{2}^{+}$, S) &0.52959$\pm$0.724$\times$10$^{-5}$ &0.11024$\pm$0.00002 & - & - \\
		($\frac{3}{2}^{+}$, S) &0.66188$\pm$0.00122 &-0.55829$\pm$0.00342 &0.71999$\pm$0.00323 &-1.04649$\pm$0.01023\\
		
		\noalign{\smallskip}
		($\frac{3}{2}^{-}$, P) &0.48958$\pm$0.00059 &-0.46751$\pm$0.00214 &- &- \\
		($\frac{5}{2}^{-}$, P) &0.57259$\pm$0.03934 &-1.11653$\pm$0.16085 &0.93538$\pm$0.13916 &-3.03894$\pm$0.61899 \\
		
		\noalign{\smallskip}
		($\frac{5}{2}^{+}$, D) &0.40653$\pm$0.00057 &-0.75873$\pm$0.00284 &- &-\\
		($\frac{7}{2}^{+}$, D) &0.45955$\pm$0.03583 &-1.31545$\pm$0.19663 &0.91999$\pm$0.18742 &-4.32966$\pm$1.10470\\
		
		\hline\noalign{\smallskip}

	\end{tabular*}
\end{table}

Table \ref{tab:tableb} shows the calculated values of $\beta_{0} $  and $\beta $ for $\Lambda$, $\Sigma$, $\Xi$, and $\Omega$ for the Regge trajectories of \textit{S, P,} and\textit{D} states. 
Hence with the aid of these extracted Regge parameters we can estimate the radial and orbital excited states of all the light strange baryons for natural and unnatural parity states. The calculated results are summarized in tables \ref{tab:table6}-\ref{tab:table9}.

\begin{table}
	\centering
	\caption{Masses of excited states of $\Lambda$ baryon in $(n,M^{2})$ plane. The masses in the boldface are taken as input from Ref. \cite{Zyla2020} (in MeV).}\label{tab:table6}
	\begin{tabular*}{140mm}{c@{\extracolsep{\fill}}cccccccccccccccc}
		\hline\noalign{\smallskip}
		&	\textit{$N^{2S+1}L_{J}$}  & Present &\cite{Menapara}& \cite{Metsch} & \cite{Capstick} &\cite{Y. Chen} & \cite{Santopinto2015} \\
		\hline\noalign{\smallskip}
		(S=1/2)& $1^{2}S_{1/2}$ &1115.68$\pm$0.006&1115 &1108 &1115 &1113 &1116\\
		& $2^{2}S_{1/2}$ &\textbf{1600$\pm$0.00}&1592 &1677 &1680 &1606 &1518\\
		& $3^{2}S_{1/2}$ &1968.57$\pm$0.01&1885 &1747 &1830 &1880 &1955\\
		& $4^{2}S_{1/2}$ &2278.27$\pm$0.01&2202 &2077 &2010 &2173 &\\
		& $5^{2}S_{1/2}$ &2550.64$\pm$0.01&2540 &2132 &2120\\
		& $6^{2}S_{1/2}$ &2796.61$\pm$0.01\\
		
		\noalign{\smallskip}\hline\noalign{\smallskip}    
		
		(S=1/2)& $1^{2}P_{3/2}$ &1551.23$\pm$0.43&1534 &1508 &1545 &1560 &1650\\
		& $2^{2}P_{3/2}$ &\textbf{1690$\pm$0.00}&1819 &1775 &1770 &1859 &1854\\
		& $3^{2}P_{3/2}$ &1818.21$\pm$3.33&2131 &2147 &2185\\
		& $4^{2}P_{3/2}$ &1937.96$\pm$3.41&2464 &2313\\
		& $5^{2}P_{3/2}$ &2050.72$\pm$3.50\\

		\noalign{\smallskip}\hline\noalign{\smallskip}    
		
		(S=1/2) & $1^{2}D_{5/2}$ &1888.09$\pm$0.50&1746 &1834 &1890 &1839 &1896\\
		& $2^{2}D_{5/2}$ &\textbf{2090$\pm$0.00}&2051 &2078 &2115 &2103\\
		& $3^{2}D_{5/2}$ &2273.39$\pm$3.27&2378\\
		& $4^{2}D_{5/2}$ &2443.05$\pm$3.37\\
		& $5^{2}D_{5/2}$ &2601.67$\pm$3.48\\
		
	\hline
	\end{tabular*}
\end{table}

\begin{table}
	\caption{Masses of excited states of $\Sigma$ baryon in $(n,M^{2})$ plane. The masses in the boldface are taken as input from Ref. \cite{Faustov} (in MeV).}\label{tab:table7}
	\begin{tabular*}{180mm}{c@{\extracolsep{\fill}}cccccccccccccccc}
		\hline\noalign{\smallskip}
		&	\textit{$N^{2S+1}L_{J}$}&\multicolumn{3}{c}{Present}& & Others	\\
		\hline\noalign{\smallskip}
		& &$\Sigma^{+}$	  & $\Sigma^{0}$ & $\Sigma^{-}$ & \cite{Menapara} & \cite{Metsch} & \cite{Capstick} &\cite{Y. Chen} 	\\
		
		\hline\noalign{\smallskip}
		(S=1/2)& $1^{2}S_{1/2}$ &1189.37$\pm$0.07&1192.64$\pm$0.024&1197.45$\pm$0.03&1193&1190&1190&1192\\
		& $2^{2}S_{1/2}$ &\textbf{1711$\pm$0.00}&\textbf{1711$\pm$0.00}&\textbf{1711$\pm$0.00}&1643&1760&1720&1664\\
		& $3^{2}S_{1/2}$ &2107.24$\pm$0.12&2105.39$\pm$0.04&2102.65$\pm$0.05&2083&1947&1915&1924\\
		& $4^{2}S_{1/2}$ &2439.95$\pm$0.14&2436.76$\pm$0.05&2432.03$\pm$0.06&2560&2098&2030&2069\\
		& $5^{2}S_{1/2}$ &2732.45$\pm$0.15&2728.17$\pm$0/05&2721.84$\pm$0.07&3067\\
		& $6^{2}S_{1/2}$ &2996.53$\pm$0.16&2991.33$\pm$0.06&2983.63$\pm$0.07\\
		
		(S=3/2)& $1^{4}S_{3/2}$ &1382.83$\pm$0.34&1383.70$\pm$1.00&1387.20$\pm$0.50&1384&1411&1370&1383\\
		& $2^{4}S_{3/2}$ &\textbf{1862$\pm$0.00}&\textbf{1862$\pm$0.00}&\textbf{1862$\pm$0.00}&1827&1896&1920&1868\\
		& $3^{4}S_{3/2}$ &2240.95$\pm$0.75&2240.42$\pm$2.22&2238.25$\pm$1.12&2229&2044&2030&2039\\
		& $4^{4}S_{3/2}$ &2564.51$\pm$0.83&2563.57$\pm$2.44&2559.78$\pm$1.22&2675&2112&2115&2122\\
		& $5^{4}S_{3/2}$ &2851.58$\pm$0.90&2850.32$\pm$2.65&2845.21$\pm$1.33&3159\\
		& $6^{4}S_{3/2}$ &3112.29$\pm$0.97&3110.74$\pm$2.86&3104.50$\pm$1.44\\   
		
		\noalign{\smallskip}\hline\noalign{\smallskip}    
		
		(S=1/2)& $1^{2}P_{3/2}$ &1605.05$\pm$0.42&1607.47$\pm$0.42&1611.04$\pm$0.41&1698&1669&1655&1698\\
		& $2^{2}P_{3/2}$ &\textbf{2175$\pm$0.00}&\textbf{2175$\pm$0.00}&\textbf{2175$\pm$0.00}&2099&1781&1755&1802\\
		& $3^{2}P_{3/2}$ &2623.94$\pm$0.91&2622.46$\pm$0.92&2620.27$\pm$0.90&2545&2203&2200\\
		& $4^{2}P_{3/2}$ &3006.58$\pm$1.01&3003.99$\pm$1.01&3000.16$\pm$0.98&3027\\
		& $5^{2}P_{3/2}$ &3345.74$\pm$1.09&3342.25$\pm$1.10&3337.09$\pm$1.07&3541\\
		
		(S=3/2) & $1^{4}P_{5/2}$ &1770.13$\pm$17.52&1770.81$\pm$17.53&1773.55$\pm$17.49&1680&1770&1755&1743\\
		& $2^{4}P_{5/2}$ &\textbf{2214$\pm$0.00}&\textbf{2214$\pm$0.00}&\textbf{2214$\pm$0.00}&2076&2174&2205\\
		& $3^{4}P_{5/2}$ &2582.68$\pm$51.46&2582.21$\pm$51.58&2580.33$\pm$51.81&2516\\
		& $4^{4}P_{5/2}$ &2904.94$\pm$55.38&2904.11$\pm$55.49&2900.76$\pm$55.73&2994\\
		& $5^{4}P_{5/2}$ &3194.86$\pm$59.41&3193.73$\pm$59.53&3189.16$\pm$59.75&3502\\
		
		\noalign{\smallskip}\hline\noalign{\smallskip}    
		
		(S=1/2) & $1^{2}D_{5/2}$ &1933.33$\pm$0.49&1935.34$\pm$0.49&1938.31$\pm$0.49&1998&1956&1995&1949\\
		& $2^{2}D_{5/2}$ &\textbf{2459$\pm$0.00}&\textbf{2459$\pm$0.00}&\textbf{2459$\pm$0.00}&2432&2071&2095&2062\\
		& $3^{2}D_{5/2}$ &2890.60$\pm$1.34&2889.26$\pm$1.34&2887.27$\pm$1.35&2905& & &2154\\
		& $4^{2}D_{5/2}$ &3265.66$\pm$1.45&3263.27$\pm$1.45&3259.75$\pm$1.46&3410\\
		& $5^{2}D_{5/2}$ &3601.86$\pm$1.56&3598.62$\pm$1.56&3593.83$\pm$1.57\\
		
		(S=3/2) & $1^{4}D_{7/2}$ &2086.75$\pm$21.02&2087.33$\pm$21.02&2089.65$\pm$20.99&1962&2070&2060&2002\\
		& $2^{4}D_{7/2}$ &\textbf{2470$\pm$0.00}&\textbf{2470$\pm$0.00}&\textbf{2470$\pm$0.00}&2390&2161&2125&2106\\
		& $3^{4}D_{7/2}$ &2801.30$\pm$81.96&2800.87$\pm$82.11&2799.14$\pm$82.54&2855\\
		& $4^{4}D_{7/2}$ &3097.36$\pm$86.59&3096.58$\pm$86.73&3093.45$\pm$78.16&3353\\
		& $5^{4}D_{7/2}$ &3367.50$\pm$91.52&3366.42$\pm$91.67&3362.09$\pm$92.08\\
	\hline
	\end{tabular*}
\end{table}

\begin{table}
	\centering
	\caption{Masses of excited states of $\Xi$ baryon in $(n,M^{2})$ plane. The masses in the boldface are taken as input from Ref. \cite{Faustov} (in MeV).}\label{tab:table8}
	\begin{tabular*}{170mm}{c@{\extracolsep{\fill}}cccccccccccccccc}
		\hline\noalign{\smallskip}
		&	\textit{$N^{2S+1}L_{J}$}&\multicolumn{2}{c}{Present}& & Others	\\
		\hline\noalign{\smallskip}
		& &$\Xi^{0}$ & $\Xi^{-}$ & \cite{Menapara} & \cite{Metsch}&\cite{Capstick} &\cite{Y. Chen} 	\\
		
		\hline\noalign{\smallskip}
		(S=1/2)& $1^{2}S_{1/2}$ &1291.80$\pm$0.01&1293.06$\pm$0.01&1322&1310&1305&1317\\
		& $2^{2}S_{1/2}$ &\textbf{1886$\pm$0.00}&\textbf{1886$\pm$0.00}&1884&1876&1840&1750\\
		& $3^{2}S_{1/2}$ &2333.51$\pm$0.01&2332.81$\pm$0.01&2361&2131&2100&2054\\
		& $4^{2}S_{1/2}$ &2708.04$\pm$0.01&2706.84$\pm$0.02&2935&2215&2150&2149\\
		& $5^{2}S_{1/2}$ &3036.73$\pm$0.02&3035.12$\pm$0.02&3591& &2345\\
		& $6^{2}S_{1/2}$ &3333.17$\pm$0.02&3331.21$\pm$0.02\\
		
		(S=3/2)& $1^{4}S_{3/2}$ &1534.38$\pm$0.91&1530.76$\pm$1.55&1531&1539&1505&1526\\
		& $2^{4}S_{3/2}$ &\textbf{1966$\pm$0.00}&\textbf{1966$\pm$0.00}&1971&1988&2045&1952\\
		& $3^{4}S_{3/2}$ &2318.62$\pm$2.41&2321.01$\pm$4.07&2457&2170&2165&2114\\
		& $4^{4}S_{3/2}$ &2624.28$\pm$2.62&2628.50$\pm$4.42&3029&2257&2230&2218\\
		& $5^{4}S_{3/2}$ &2897.87$\pm$2.82&2903.61$\pm$4.71&3679\\
		& $6^{4}S_{3/2}$ &3147.78$\pm$3.02&3154.82$\pm$5.11\\   
		
		\noalign{\smallskip}\hline\noalign{\smallskip}    
		
		(S=1/2)& $1^{2}P_{3/2}$ &1731.32$\pm$0.71&1732.26$\pm$0.71&1871&1780&1785&1801\\
		& $2^{2}P_{3/2}$ &\textbf{2245$\pm$0.00}&\textbf{2245$\pm$0.00}&2337&1924&1895&1976\\
		& $3^{2}P_{3/2}$ &2661.31$\pm$1.80&2660.71$\pm$1.80&2894&2353&2330\\
		& $4^{2}P_{3/2}$ &3020.78$\pm$1.95&3019.71$\pm$1.96&3532\\
		& $5^{2}P_{3/2}$ &3341.81$\pm$2.11&3340.36$\pm$2.12\\
		
		(S=3/2) & $1^{4}P_{5/2}$ &1922.61$\pm$31.21&1919.72$\pm$31.27&1859&1955&1900&1917\\
		& $2^{4}P_{5/2}$ &\textbf{2333$\pm$0.00}&\textbf{2333$\pm$0.00}&2318&2292&2345\\
		& $3^{4}P_{5/2}$ &2681.30$\pm$105.97&2683.37$\pm$105.36&2865&2438\\
		& $4^{4}P_{5/2}$ &2989.28$\pm$112.94&2992.99$\pm$112.34&3494\\
		& $5^{4}P_{5/2}$ &3268.37$\pm$120.23&3273.47$\pm$119.64\\
		
		\noalign{\smallskip}\hline\noalign{\smallskip}    
		
		(S=1/2) & $1^{2}D_{5/2}$ &2079.95$\pm$0.83&2080.74$\pm$0.84&2234&2013&2045&1959\\
		& $2^{2}D_{5/2}$ &\textbf{2605$\pm$0.00}&\textbf{2605$\pm$0.00}&2771&2197&2230&2170\\
		& $3^{2}D_{5/2}$ &3040.70$\pm$2.42&3040.16$\pm$2.45&3391&2279&2240&2239\\
		& $4^{2}D_{5/2}$ &3421.36$\pm$2.61&3420.40$\pm$2.64&\\
		& $5^{2}D_{5/2}$ &3763.71$\pm$2.80&3762.40$\pm$2.84\\
		
		(S=3/2) & $1^{4}D_{7/2}$ &2244.67$\pm$37.79&2242.19$\pm$56.72&2203&2169&2180&2074\\
		& $2^{4}D_{7/2}$ &\textbf{2686$\pm$0.00}&\textbf{2686$\pm$0.00}&2729&2289&2240&2189\\
		& $3^{4}D_{7/2}$ &3064.42$\pm$138.35&3066.23$\pm$137.53&3336\\
		& $4^{4}D_{7/2}$ &3400.98$\pm$146.74&3404.26$\pm$145.92\\
		& $5^{4}D_{7/2}$ &3707.12$\pm$155.60&3711.02$\pm$154.78\\
		
	\hline
	\end{tabular*}
\end{table}

\begin{table}
	\centering
	\caption{Masses of excited states of $\Omega$ baryon in $(n,M^{2})$ plane. The masses in the boldface are taken as input from Ref \cite{ChandniOmega} (in MeV).}\label{tab:table9}
	\begin{tabular*}{130mm}{c@{\extracolsep{\fill}}cccccccccccccccc}
		\hline\noalign{\smallskip}
		& \textit{$N^{2S+1}L_{J}$} & Present & \cite{Faustov} &\cite{Y. Chen} &\cite{Metsch} & \cite{Capstick}\\
		\hline\noalign{\smallskip}
		(S=3/2)& $1^{4}S_{3/2}$ &1685.93$\pm$1.85&1678&1673& &1635\\
		& $2^{4}S_{3/2}$ &\textbf{2057$\pm$0.00}&2173&2078&2177&2165\\
		& $3^{4}S_{3/2}$ &2370.68$\pm$6.11\\
		& $4^{4}S_{3/2}$ &2647.46$\pm$6.52\\
		& $5^{4}S_{3/2}$ &2897.92$\pm$6.95\\
		& $6^{4}S_{3/2}$ &3128.39$\pm$7.38\\   
		
		\noalign{\smallskip}\hline\noalign{\smallskip}

		(S=3/2) & $1^{4}P_{5/2}$ &2077.97$\pm$38.27&2653 & &2528&2490\\
		& $2^{4}P_{5/2}$ &\textbf{2321$\pm$0.00}& & &2534\\
		& $3^{4}P_{5/2}$ &2540.89$\pm$229.51\\
		& $4^{4}P_{5/2}$ &2743.21$\pm$237.03\\
		& $5^{4}P_{5/2}$ &2931.60$\pm$245.53\\
		
		\noalign{\smallskip}\hline\noalign{\smallskip}

		(S=3/2) & $1^{4}D_{7/2}$ &2406.97$\pm$46.70&2369&2205&2292&2295\\
		& $2^{4}D_{7/2}$ &\textbf{2623$\pm$0.00}\\
		& $3^{4}D_{7/2}$ &2822.16$\pm$357.64\\
		& $4^{4}D_{7/2}$ &3008.99$\pm$365.72\\
		& $5^{4}D_{7/2}$ &3184.50$\pm$375.18\\
		
	\hline
	\end{tabular*}
\end{table}

\subsection{Other states in the ($J,M^{2}$) plane}

So far, we have used conventional formulae to compute the masses of light strange baryons for natural and unnatural parity states. After the successful implementation of this model, now in this section, we try to obtain the remaining other states in the ($J,M^{2}$) plane by using the same method. Since we have calculated $1^{2}P_{\frac{3}{2}}$ and $1^{4}P_{\frac{5}{2}}$ states earlier, now here we firstly calculate the other three $1P$ states i.e., $1^{2}P_{\frac{1}{2}}$, $1^{4}P_{\frac{1}{2}}$, and $1^{4}P_{\frac{3}{2}}$ by using the Eq. (\ref{eq:8}). For $\Xi$ baryon we put $i=d$, $j=s$, $q=u$, and $k=d$ in Eq. (\ref{eq:8}) we get same relation (\ref{eq:9}) in terms of light baryon masses which is expressed as, 
\begin{eqnarray}
	\label{eq:16} 
	\left[(M_{n}+M_{\Xi})^{2} - 4M^{2}_{\Lambda}\right] 
	= \sqrt{(4M^{2}_{\Lambda}-M^{2}_{n}-M^{2}_{\Xi})^{2} - 4M^{2}_{n}M^{2}_{\Xi}} ;
\end{eqnarray}
Here also, after putting the masses in above equation we get masses for $1^{2}P_{\frac{1}{2}}$, $1^{4}P_{\frac{1}{2}}$, and $1^{4}P_{\frac{3}{2}}$ states. Similarly we can determine the other $1P$ states masses for $\Omega$ baryon as well.
Once we have calculated the $1P$ states, we extract the Regge slopes  for all the light strange baryons in the $(J,M^{2})$ plane as we have done in previous section. Again using Eq. (\ref{eq:1}) we have,

\begin{equation}
	\label{eq:17} 
	\nonumber
	M_{J+1} = \sqrt{M_{J}^{2}+\frac{1}{\alpha^{'}}} .
\end{equation}

Excited state masses can be obtained by using the above relation.  Tables \ref{tab:table10} - \ref{tab:table13} shows our calculated results for the remaining other states for light strange baryons. We compared our estimated masses with experimental data and other theoretical studies. 

\begin{table}
	\centering
	\caption{Masses of other excited  states of $\Lambda$ baryon in the $(J,M^{2})$ plane. The numbers in the boldface are the experimental values taken as the input \cite{Zyla2020} (in MeV).}\label{tab:table10}
	\begin{tabular*}{125mm}{c@{\extracolsep{\fill}}cccccccccccccccc}
		\hline\noalign{\smallskip}
		\textit{$N^{2S+1}L_{J}$}& Present & PDG \cite{Zyla2020} & \cite{Menapara} & \cite{Faustov} & \cite{Metsch}& \cite{Capstick} & \cite{Y. Chen} \\
		\hline\noalign{\smallskip}
		$1^{2}P_{\frac{1}{2}}$ & \textbf{1670} &1670 &1546 &1406& 1524&1550 &1559\\
		$1^{2}D_{\frac{3}{2}}$ & 1861 &1890 &1769 &1854& 1823&1900&1836\\
		$1^{2}F_{\frac{5}{2}}$ & 2034 &2082$\pm$13 &2005 &2136&2080 &2180\\
		$1^{2}G_{\frac{7}{2}}$ & 2193 & &2253\\
		$1^{2}H_{\frac{9}{2}}$ & 2342 \\
		
	\hline
	\end{tabular*}
\end{table}

\begin{table}
	\centering
	\caption{Masses of other excited  states of $\Sigma$ baryon in the $(J,M^{2})$ plane. The numbers in the boldface are taken as the input (in MeV).}\label{tab:table11}
	\begin{tabular*}{115mm}{c@{\extracolsep{\fill}}cccccccccccccccc}
		\hline\noalign{\smallskip}
		\textit{$N^{2S+1}L_{J}$}& Present &PDG\cite{Zyla2020}& \cite{Faustov} &\cite{Capstick}&\cite{Y. Chen} &\cite{Metsch}   \\
		\hline\noalign{\smallskip}
		$1^{2}P_{\frac{1}{2}}$ & \textbf{1620} \cite{Zyla2020}& &1620&1630 &1657 &1628\\
		$1^{4}P_{\frac{1}{2}}$ & \textbf{1750} \cite{Zyla2020} &&1693 &1675 &1746 &1771\\
		$1^{4}P_{\frac{3}{2}}$ & \textbf{1709} \cite{Menapara} &&1731 &1750 &1790 &1728\\
		$1^{2}D_{\frac{3}{2}}$ & 1816 &&2025 &1970 &1947 &1961\\
		$1^{4}D_{\frac{3}{2}}$ & 2062 &2090&2076 &2010 &1993 &2011\\
		$1^{4}D_{\frac{5}{2}}$ & 1997 &&2062 &2030 &2028 &2027\\
		$1^{2}F_{\frac{5}{2}}$ & 1993 &&2347 &2250 & & 2226\\
		$1^{4}F_{\frac{5}{2}}$ & 2333 &&\\
		$1^{4}F_{\frac{7}{2}}$ & 2249 &2250&2349 & & &2285\\
		$1^{2}G_{\frac{7}{2}}$ & 2155&\\
		$1^{4}G_{\frac{7}{2}}$ & 2575&\\
		$1^{4}G_{\frac{9}{2}}$ & 2475&2455\\
	\hline
	\end{tabular*}
\end{table}

\begin{table}
	\centering
	\caption{Masses of other excited  states of $\Xi$ baryon in the $(J,M^{2})$ plane. (in MeV).}\label{tab:table12}
	\begin{tabular*}{115mm}{c@{\extracolsep{\fill}}cccccccccccccccc}
		\hline\noalign{\smallskip}
		\textit{$N^{2S+1}L_{J}$}& Present&PDG \cite{Zyla2020} & \cite{Faustov} &\cite{Capstick} &\cite{Metsch} &\cite{Y. Chen}    \\	
		\hline\noalign{\smallskip}
		$1^{2}P_{\frac{1}{2}}$ & 1810 &&1682 &1755 &1770 &1772\\
		$1^{4}P_{\frac{1}{2}}$ & 1890 &&1758 &1810 &1922 &1894\\
		$1^{4}P_{\frac{3}{2}}$ & 1825 &1823$\pm$5&1798 &1880 &1873 &1918\\
		$1^{2}D_{\frac{3}{2}}$ & 2155 &&2100 &2065 &2076 &1970\\
		$1^{4}D_{\frac{3}{2}}$ & 2204 &&2121 &2115 &2128 &2065\\
		$1^{4}D_{\frac{5}{2}}$ & 2115 &2120&2147 &2165 &2141 &2102\\
		$1^{2}F_{\frac{5}{2}}$ & 2318 &&2411 &2350 &2409 &\\
		$1^{4}F_{\frac{5}{2}}$ & 2478 && &2385 &2425\\
		$1^{4}F_{\frac{7}{2}}$ & 2370 &2370&2474 &2425\\
		$1^{2}G_{\frac{7}{2}}$ & 2470\\
		$1^{4}G_{\frac{7}{2}}$ & 2725\\
		$1^{4}G_{\frac{9}{2}}$ & 2600\\
	\hline
	\end{tabular*}
\end{table}

\begin{table}
	\centering
	\caption{Masses of other excited states of $\Omega$ baryon in the $(J,M^{2})$ plane (MeV).}\label{tab:table13}
	\begin{tabular*}{115mm}{c@{\extracolsep{\fill}}cccccccccccccccc}
		\hline\noalign{\smallskip}
		\textit{$N^{2S+1}L_{J}$}& Present &PDG \cite{Zyla2020}&  \cite{Faustov} &\cite{Capstick} &\cite{Metsch} & \cite{Y. Chen} \\	
		\hline\noalign{\smallskip}
		$1^{4}P_{\frac{1}{2}}$ & 2030&2012.4$\pm$0.9 &2463 &2410 &2456 &\\
		$1^{4}P_{\frac{3}{2}}$ & 1941&&2537 &2440 &2446 &\\
		
		$1^{4}D_{\frac{3}{2}}$ & 2338&2380&2332 &2345 &2287 &2263\\
		$1^{4}D_{\frac{5}{2}}$ & 2249&2252$\pm$9& &2345 &2312 &2260\\
		
		$1^{4}F_{\frac{5}{2}}$ & 2610 && & &2617\\
		$1^{4}F_{\frac{7}{2}}$ & 2519 &&2599 & &2531\\
		
		$1^{4}G_{\frac{7}{2}}$ & 2856&\\
		$1^{4}G_{\frac{9}{2}}$ & 2763&\\
	\hline
	\end{tabular*}
\end{table}

\section{Results and Discussion}

In the present work, an attempt has been made to obtain the mass spectra of hyperons under the methodology of Regge Phenomenology. Regge slopes  were calculated in the $(J,M^{2})$ plane. With the aid of these Regge slopes, the masses of the orbitally excited  baryons were estimated for both natural and unnatural parity states. After that, the Regge slopes and intercepts were extracted for each Regge line in the $(n,M^{2})$ plane, and with the help of these parameters mass spectra of light baryons were obtained successfully. Tables \ref{tab:table2}-\ref{tab:table5} and \ref{tab:table6}-\ref{tab:table9} summarizes the calculated masses in the $(J,M^{2})$ and  $(n,M^{2})$ planes respectively, for natural and unnatural parity states along with the other theoretical outcomes and the experimental observations where available. Also, our estimated masses for remaining other states in the $(J,M^{2})$ plane are shown in tables \ref{tab:table10}-\ref{tab:table13}.


\begin{enumerate}
	
	\item For $\Lambda$ baryon the four star and three star status states $\Lambda(1520)$, $\Lambda(1820)$,  $\Lambda(1890)$,  $\Lambda(2080)$, $\Lambda(2100)$, and $\Lambda(2350)$ mentioned in the PDG \cite{Zyla2020} with well established quantum numbers are in agreement with our calculated results. Some states shows a slight discrepancies of mass difference of around 60-80 MeV. We confirmed the $J^{P}$ values of these states in the present work (see tables \ref{tab:table2} and \ref{tab:table10}). We compared our results with the predictions of other theoretical models also and our estimated masses are consistent with them as shown in Tables \ref{tab:table2}, \ref{tab:table6}, and \ref{tab:table10}. The obtained masses for the states $1^{2}S_{\frac{1}{2}}$ and $1^{2}P_{\frac{3}{2}}$ are ,close to the results of Refs. \cite{Faustov,Menapara,Capstick} with a mass difference of 1-18 MeV. Also the results of further higher excited states $1^{2}D_{\frac{5}{2}}$ and $1^{2}F_{\frac{7}{2}}$ are in good agreement with the results of Refs. \cite{Capstick,Santopinto2015} with a mass difference of 0-24 MeV and slightly vary with the results of other references. Also, the calculated results in the ($n,M^{2}$) plane are in  consistent with the outcomes of other theoretical models.\\
		
	\item The calculated results in the $(J,M^{2})$ and $(n,M^{2})$ plane for $\Sigma$ baryon are summerized in Tables \ref{tab:table3}, \ref{tab:table7}, and \ref{tab:table11}. For the case of $\Sigma$ baryon the masses of well established states; $\Sigma(1670)$, $\Sigma(1775)$, $\Sigma(1915)$, $\Sigma(2030)$, and $\Sigma(2080)$ closely matches with the masses of our estimated results, and our model confirmed the spin parity of these states. The $\Sigma(2455)$, $\Sigma(2620)$, and $\Sigma(2250)$ states are mentioned in PDG with two and three star but their $J^{P}$ values are still unknown. The $\Sigma(2250)$ is very close to our predicted mass 2249 MeV having a mass difference of 1 MeV only (see table \ref{tab:table11}). So we predicted this state to be $1^{4}F_{\frac{7}{2}}$ having spin parity $\frac{7}{2}^{-}$. The states $\Sigma(2455)$ and $\Sigma(2620)$ are fairly close to our estimated masses 2475 MeV and 2609.37$\pm$23.77 with a mass difference of 20 MeV and 11 MeV respectively. So we can say that these two states may belong to $1G$ state having $J^{P}= \frac{9}{2}^{+}$ and $\frac{11}{2}^{+}$ (see tables \ref{tab:table3} and \ref{tab:table11}).
	Also, the calculated masses for low lying resonance fairly matches with the results of other theoretical predictions but a wide range of mass difference is shown for higher excited states. Our obtained masses in the ($J,M^{2}$) plane for $1S$, $1P$, and $1D$ are close to the results of Refs. \cite{Metsch,Capstick} with a mass difference of few MeV and shows slightly large mass difference with the results of other theoretical predictions. Similarly, we compared our calculated results in the ($n,M^{2}$) plane with the results of Refs. \cite{Menapara,Metsch,Capstick,Y. Chen}. The masses for low lying states are in good agreement with the results of other theoretical predictions but shows slight large mass difference for higher excited states.\\
	
	Our calculated ground-state masses of $\Xi$ baryon as 1291.80$\pm$0.01 MeV ($\Xi^{0}$), 1293.06$\pm$0.01 MeV ($\Xi^{-}$) for $J^{P}= \frac{1}{2}^{+}$ is in good agreement with the experimental mass having a mass difference of around 24 MeV and for $J^{P}= \frac{3}{2}^{+}$ our obtained masses 1534.38$\pm$0.91 MeV ($\Xi^{0*}$), 1530.76$\pm$1.55 MeV ($\Xi^{-*}$) are very close to the experimental masses with a mass difference of only 2-5 MeV. They  also matches very well with the predictions of other theoretical models (see table \ref{tab:table4})
	
Experimentally, very few states are confirmed with spin-parity in the $\Xi$ family. The $\Xi(1820)$ is the only negative parity state assigned with $J^{P}=\frac{3}{2}^{-}$ in PDG having mass 1823 MeV which is very close our predicted mass 1825 MeV with a slight difference of 2 MeV only for  $1^{4}P_{\frac{3}{2}}$ state. The $\Xi(2030)$ is assigned with angular momentum having value $\frac{5}{2}$, parity of this state is not confirmed yet. Here we predicted this state with positive parity having mass 2079.95$\pm$0.83 MeV belongs to $1D$ state having $J^{P} = \frac{5}{2}^{+}$. The $J^{P}$ value of three stared $\Xi(1950)$ state is still not confirmed in PDG. Our predicted mass 1922.61$\pm$31.21 MeV for $1^{4}P_{\frac{5}{2}}$ state is near to 1950$\pm$15 Mev with a mass difference of 28 MeV. Also our predicted mass lies in the range of calculated experimental error. So, we predicted the spin-parity of this state to be $\frac{5}{2}^{-}$ for S = 3/2.  The two starred state $\Xi(2250)$ is close to our estimated mass 2244.67$\pm$37.79 MeV with a mass difference of 5 MeV, so we predicted this state as $1D$ state with $J^{P} = \frac{7}{2}^{+}$ for $S=3/2$. The $\Xi(2370)$ state with two star matches exactly with our predicted mass 2370 MeV (see table \ref{tab:table12}) and also it is very close to 2378.01$\pm$0.89 MeV (see table \ref{tab:table4}). So this state may belongs to either $1^{2}F_{\frac{7}{2}}$ or $1^{4}F_{\frac{7}{2}}$ having $J^{P} = \frac{7}{2}^{-}$. Also, one more state $\Xi(2120)$ with one star is found to be very close to our mass 2115 MeV with a mass difference of 5 MeV (see table \ref{tab:table12}). Hence we can say that, this may belongs to $1D$ state with $J^{P}=\frac{5}{2}^{+}$ for $S=3/2$. We compared our calculated results with various theoretical approaches and they are in  consistent with the prediction of them. The calculated masses for $1P$, $1D$, and $1F$ states are very close to the predicted masses of Ref. \cite{Faustov} with a mass difference of few MeV. Our results are also in consistent with the predictions of \cite{Metsch,Capstick} and shows slightly large mass difference. Similarly, in the ($n,M^{2}$) plane our predicted masses for low lying states are in accordance with the results of other theoretical approaches but for higher excited states variation in mass differences observed with other theoretical results. \\

\item The estimated results for $\Omega$ baryon in the $(J,M^{2})$ and $(n,M^{2})$ planes are summerized in tables \ref{tab:table5}, \ref{tab:table9}, and \ref{tab:table13}.  Experimentally observed ground state of $\Omega$ baryon with known spin parity $\frac{3}{2}^{+}$ having mass 1672.45$\pm$0.29 MeV is found to be very close to our estimated mass 1685.93$\pm$1.85 MeV with a slight mass difference of 13.5 Mev (see table \ref{tab:table5}). Four new resonances states of $\Omega$ baryon ; $\Omega(2012)$, $\Omega(2250)$, $\Omega(2380)$, and $\Omega(2470)$ are experimentally observed. 
Their spin-parity are still not confirmed. The three-star state $\Omega(2012)$ is still a mystery. Because it is so close to the $\Xi(1530)K$ threshold, a recent study suggested that this state is of molecular nature \cite{M. V. Polyakov2019}. Other studies \cite{M. Pavon2018,Y. H. Lin2018} indicate that the current information is insufficient to consider it as a molecular state.  However in the present study this state is close to the predicted mass 2030 MeV vary by 18 MeV (see table \ref{tab:table13}), so this state may belong to $1^{4}P_{\frac{1}{2}}$ having $J^{P}=\frac{1}{2}^{-}$. The $\Omega(2250)$ state having mass 2252 MeV is very close to our calculated mass 2249 MeV with a slight mass differences of 3 MeV, hence we can say that this state can be a good candidate of $D$-wave and identified with spin parity $J^{P}=\frac{5}{2}^{+}$ for S=3/2. The two starred state $\Omega(2380)$ in PDG is close to our predicted mass 2406.97$\pm$46.70 MeV with a mass difference of 26 MeV and also this state is reasonably close to our calculated mass 2338 MeV with a mass difference of 42 MeV. So this state belongs to $D$-wave and may have spin-parity $\frac{7}{2}^{+}$ or $\frac{3}{2}^{+}$.
The $\Omega(2470)$ state is not identified in this work.	We compared our results with other theoretical predictions as well and they are in good agreement with the assumptions of Refs. \cite{Faustov,Capstick,Metsch,Y. Chen}. Also we observed a wide range of mass difference for $\Omega(2012)$ state with other theoretical results.
	
\end{enumerate}

\section{Conclusion} 

Regge phenomenology has been effectively implemented to study the mass-spectra of light baryons with strangeness -1,-2, and -3. Here our aim of determining the spin parity of experimentally observed states : $\Sigma(2250)$, $\Sigma(2455)$, $\Sigma(2620)$, 
$\Xi(1950)$, $\Xi(2030)$, $\Xi(2120)$, $\Xi(2250)$, $\Xi(2370)$, $\Omega(2012)$, $\Omega(2250)$, and $\Omega(2380)$ is accomplished.
Also we confirmed the spin parity of many experimentally well established states of $\Lambda$ and $\Sigma$ baryons.
Here the error propogation is incorporated and we have calculated the experimental error in the obtained masses, wherever possible and  the experimental masses are taken as inputs. Our obtained results are very close to the experimental masses with a mass difference of few MeV. Also, the experimentally established masses lies within the error range of the calculated masses. 
It is evident that the low-lying resonance masses are in good agreement with other theoretical predictions but for higher excited states we see a wide range of mass  difference. This possibly because of the fact that not a single model exactly predicts the spin-parity assignments, and also there is no experimental evidence for the states. As a result, our work with a huge number of spin-parity predictions could be useful for future facilities like $\overline{\mbox{\sffamily P}}${\sffamily ANDA}, which is intended to explore light strange baryons in depth.




\begin{thebibliography}{90}


\bibitem{ALICE2019}
S. Acharya \textit{et al.} (ALICE Collaboration), Phys. Rev. Lett.
\textbf{123} (2019)  112002. 
\bibitem{BESIII2019}
M. Ablikim \textit{et al.} (BESIII Collaboration), Phys. Rev. D \textbf{100} (2019)  051101.
\bibitem{Jefferson2005}
W. Price \textit{et al.} (CLAS Collaboration), Phys. Rev. C \textbf{71} (2005)  058201.  
\bibitem{Jlab2020}
M. Amaryam \textit{et al.} (KLF Collaboration), arXiv:
2008.08215 [nucl-ex] (2020)
\bibitem{BABAR2006}
B. Aubert \textit{et al.} (BABAR Collaboration), Phys. Rev. Lett.
\textbf{97} (2006) 112001.

\bibitem{Yelton2018}
J. Yelton et al. (Belle Collaboration) Phys. Rev. Lett. \textbf{121} (2018) 052003. 
\bibitem{Pervin}
M. Pervin and W. Roberts, Phys. Rev. C \textbf{77} (2008) 025202.   
\bibitem{Aaij2017}
R. Aaij et al. (LHCb Collaboration) Phys. Rev. Lett. \textbf{118} (2017)  182001.  
\bibitem{Yelton2018_2}
J. Yelton et al. (Belle Collaboration) Phys. Rev. D \textbf{97} (2018) 051102. 

\bibitem{Panda1}
B. Singh \textit{et al.} (PANDA Collaboration), J. Phys. G: Nucl.
Part. Phys. \textbf{46} (2019) 045001.  
\bibitem{Panda2}
B. Singh \textit{et al.} (PANDA Collaboration), Phys. Rev. D
\textbf{95} (2017)  032003.
\bibitem{Panda3}
B. Singh \textit{et al.} (PANDA Collaboration), Eur. Phys. J. A
\textbf{52} (2016) 325.  
\bibitem{Panda4}
B. Singh \textit{et al.} (PANDA Collaboration), Eur. Phys. J. A \textbf{51} (2015) 107. 

\bibitem{Panda6}
B. Singh \textit{et al.} (PANDA Collaboration), Nucl. Phys. A 954 (2016) 323-340.

\bibitem{Barucca2021_1}
G. Barucca \textit{et al.} ($\overline{\mbox{\sffamily P}}${\sffamily ANDA}), Eur. Phys. J. A \textbf{57} (2021) 184.  
  
\bibitem{Panda5}
G. Barruca\textit{et al.} (PANDA Collaboration), Eur. Phys. J. A
\textbf{55} (2019) 42; arXiv:2009.11582; arXiv:2101.11877; arXiv:
2012.01776.
\bibitem{Abazov}V. Abazov \textit{et. al.} (PANDA Collaboration), arXiv:2201.03852.
\bibitem{Barucca2021_2}
G. Barucca \textit{et al.} ($\overline{\mbox{\sffamily P}}${\sffamily ANDA}), Eur. Phys. J. A \textbf{57} (2021) 149.
\bibitem{Putz2020}
J.P\"{u}tz, PhD thesis, Rhesinishe Friedrich-Wilhelms-University\"{a}t Bonn (2020)
\bibitem{Zyla2020}
R.L. Workman \textit{et al.} (Particle Data Group), Prog. Theor. Exp. Phys.\textbf{ 2022} (2022) 083C01. 

\bibitem{Gell-Mann}
M. Gell-Mann, Phys. Rev. \textbf{125} (1962) 1067 ; Phys. Lett. \textbf{8} (1964) 214; S. Okubo, Prog. Theor. Phys. \textbf{27} (1962) 949; \textbf{28} (1962) 24.

\bibitem{Hua-Xing2023}Hua-Xing Chen \textit{et al.}  Rep. Prog. Phys. \textbf{86} (2023) 026201.
\bibitem{F. Gross2023}F. Gross \textit{et al.}  arXiv:2212.11107v2 [hep-ph] (2023).

\bibitem{Faustov}
R. N. Faustov and V. O. Galkin, Phys. Rev. D \textbf{92} (2015) 054005. 
\bibitem{Menapara}
C. Menapara and A.K. Rai, Chin. Phys. C \textbf{45} (2021) 063108; Chin. Phys. C \textbf{45} (2021) 023102.
\bibitem{ChandniIJMPA}C. Menapara and A.K. Rai, Int. J. Mod. Phys. A \textbf{37} (2022) 27, 2250177.
\bibitem{ChandniIJMPA2023}C. Menapara and A.K. Rai, Int. J. Mod. Phys. A (2023) (accepted).
 
\bibitem{ChandniOmega}
C. Menapara and A.K. Rai, Chin. Phys. C \textbf{46} (2022)  103102.
\bibitem{zalak}Z. Shah, K. Gandhi and A. K. Rai, Chin. Phys. C \textbf{43} (2019) 034102.

\bibitem{Metsch}
U. \textbf{L\"{o}ring}, B. Ch. Metsch, and H. R. Petry, Eur. Phys. J. A \textbf{10} (2001) 447. 
\bibitem{Xiao2021}Z.-Yu Wang
, H. A. Ahmed, and C. W. Xiao, Eur. Phys. J. C \textbf{81} (2021)  833.
\bibitem{Bijker2000}
R. Bijker, F. Iachello, and A. Leviatan, Ann. Phys. (N.Y.) \textbf{284} (2000)  89-113.
\bibitem{ELSA}H. Schmieden, EPJ Web of Conferences \textbf{199} (2019) 01017.   


\bibitem{Oudichhya}
J. Oudichhya, K. Gandhi, and A.K. Rai, Phys. Rev. D \textbf{103} (2021) 114030.  
\bibitem{Oudichhya2}J. Oudichhya, K. Gandhi, and A.K. Rai, Phys. Rev. D \textbf{104} (2021) 114027.
\bibitem{physica}
J. Oudichhya, K. Gandhi, and A.K. Rai, Phys. Scr. 97 (2022) 054001.
\bibitem{Juhiarxiv}J. Oudichhya, K. Gandhi, and A.K. Rai,  arXiv:2304.05110v1 [hep-ph].
\bibitem{Wei2008}
X. H. Guo, K. W.Wei and X. H.Wu, Phys. Rev. D {\bf 78} (2008) 056005. 


\bibitem{Chew1961}
G. F. Chew and S. C. Frautschi, Phys. Rev. Lett. \textbf{7} (1961) 394.  
\bibitem{Nambu1974}
Y. Nambu, Phys. Rev. D \textbf{10} (1974) 4262.  
\bibitem{Nambu1979}
Y. Nambu, Phys. Lett. B \textbf{80} (1979) 372. 



\bibitem{Add4}
A. B. Kaidalov, Z. Phys. C \textbf{12} (1982) 63.  
\bibitem{Add3}
L. Burakovsky and T. Goldman, Phys. Lett. B \textbf{434} (1998) 251.  
\bibitem{Add6}V.V. Dixit and L. A. P. Balazs, Phys. Rev. D \textbf{20} (1979) 816.
\bibitem{Approx}
L. Burakovsky and T. Goldman, and L. P Horwitz, Phys. Rev. D \textbf{56} (1997) 7124.  


\bibitem{Capstick}
S. Capstick and N. Isgur, Phys. Rev. D \textbf{34} (1986) 2809.  
\bibitem{Santopinto2015}
E. Santopinto and J. Ferretti, Phys. Rev. C \textbf{92} (2015) 025202. 
\bibitem{Y. Chen}
Y. Chen and B.-Q. Ma, Nucl. Phys. A \textbf{831} (2009) 1. 


\bibitem{M. V. Polyakov2019}
M. V. Polyakov, H.-D. Son, B.-D. Sun, and A. Tandogan, Phys. Lett. B \textbf{792} (2019) 315-319.  
\bibitem{M. Pavon2018}
M. Pavon Valderrama, Phys. Rev. D \textbf{98} (2018) 054009. 
\bibitem{Y. H. Lin2018}
Y. H. Lin and B. S. Zou, Phys. Rev. D \textbf{98} (2018) 056013. 


\end{thebibliography}


\end{document}